\documentclass[11pt]{article}

\usepackage[a4paper,margin=2.5cm]{geometry}

\usepackage{amsmath,amssymb,bm}
\usepackage{amsthm}

\usepackage{graphicx}
\usepackage{booktabs}
\usepackage{multirow}
\usepackage{makecell}
\usepackage{array}
\usepackage{float}
\usepackage{enumitem}
\usepackage{caption}
\usepackage{algorithm}
\usepackage{algpseudocode}
\usepackage{rotating}
\usepackage{xcolor}
\usepackage[normalem]{ulem}
\usepackage{microtype}
\usepackage{authblk}
\usepackage{comment}
\usepackage{natbib}

\usepackage[hidelinks]{hyperref}

\theoremstyle{plain}
\newtheorem{proposition}{Proposition}

\theoremstyle{remark}
\newtheorem{remark}{Remark}

\DeclareMathOperator{\argmax}{arg\,max}
\DeclareMathOperator{\argmin}{arg\,min}

\newcommand{\E}{\mathbb{E}}
\newcommand{\Var}{\operatorname{Var}}
\newcommand{\Cov}{\operatorname{Cov}}

\newcommand{\btheta}{\bm{\theta}}
\newcommand{\bgamma}{\bm{\gamma}}
\newcommand{\htheta}{\hat{\bm{\theta}}}

\newcommand{\bphi}{\bm{\phi}}
\newcommand{\bbeta}{\bm{\beta}}
\newcommand{\balpha}{\bm{\alpha}}
\newcommand{\KL}{\mathrm{KL}}

\begin{document}


\title{\texorpdfstring{
\(\bm{J}\)- and \(\bm{MJ}\)-Type Tests for Non-Nested Parametric
Survival Models with a Cure Fraction:
A Score Test Approach
}
{
J- and MJ-Type Tests for Non-Nested Parametric Survival Models with a Cure Fraction: A Score Test Approach
}}


\author[1,3]{Cynthia A. V. Tojeiro}
\author[2]{Francisco Cribari-Neto}
\author[3]{Tatiene C. Souza}
\author[3]{Tarciana L. Pereira}

\affil[1]{Institute of Mathematics and Statistics, Federal University of Goi\'as, Goi\^ania, GO, Brazil}
\affil[2]{Department of Statistics, Federal University of Pernambuco, Recife, PE, Brazil}
\affil[3]{Department of Statistics, Federal University of Para\'iba, Jo\~ao Pessoa, PB, Brazil}

\date{}

\maketitle

\begin{abstract}
We propose specification tests for discriminating among non-nested
parametric survival models with a cure fraction, focusing on models that
differ only in their baseline distributions. The proposed approach augments
the null log-likelihood with information from competing models and applies a
score test to assess whether the additional information is redundant.
Because the test relies only on restricted maximum likelihood estimates, it
avoids fitting augmented models.
For two competing models, the score statistic reduces to a quadratic form in
the sample mean of the individual log-likelihood differences. We show that
its signed square root coincides with Vuong's test statistic,
although our framework differs in three important respects: it tests the
specific null hypothesis that a given model is the true data-generating
process, it uses an unsigned statistic that extends naturally to
\(M \geq 2\) competing models, and it estimates the Kullback--Leibler bias
by parametric bootstrap.
The resulting \(MJ\) statistic combines the individual \(J\) tests to assess
the global null hypothesis that at least one candidate model is correctly
specified, while also providing a model-selection criterion.
\end{abstract}


\maketitle

\section{Introduction}\label{sec:introduction}

Parametric survival analysis requires the analyst to specify a
probability distribution for the time-to-event variable. In practice,
however, the true data-generating mechanism is rarely known, and
several plausible distributions are often entertained. The choice among
competing specifications can substantially affect parameter estimation,
predictive performance, and inferential conclusions. This issue becomes
particularly relevant in settings where a non-negligible fraction of
individuals will never experience the event of interest. In such cases,
long-term survival models, also known as cure-fraction models, provide a
natural and widely used framework by explicitly accommodating a cured
subpopulation \cite{boag1949,berk,maller1996,rodrigues,tso,rondeau2013, su2022}.

Model selection in parametric survival analysis is most commonly based
on information criteria such as the Akaike information criterion (AIC) \cite{akaike}
and the Bayesian information criterion (BIC) \cite{schwarz1978}. When the competing models 
are nested, the likelihood ratio test offers an additional formal
mechanism for model discrimination \cite{cox1974} \cite{marinho2020}. For non-nested
models, however, AIC and BIC remain the dominant tools, since they can
compare arbitrary candidate distributions. Despite their practical
appeal, these criteria provide only relative rankings: they necessarily
select a ``best'' model among the candidates, regardless of whether any
of them offers an adequate description of the data. If all candidate
models are misspecified, AIC and BIC will still favor one of them. This
limitation is particularly consequential in large samples, where even
mild but systematic departures from the assumed distributions may lead
to materially different inferential conclusions. The question of whether
the true data-generating process belongs to the candidate model set is
therefore fundamental. Addressing it requires a testing framework
capable of rejecting all competing models
simultaneously---signaling that the analyst should
consider a broader or more flexible family of distributions---a
possibility beyond the scope of information criteria and standard
likelihood-ratio tests for nested models.

The problem of testing non-nested parametric models has a long and
influential history. The pioneering contributions of Cox
\cite{cox1961,cox1962} introduced likelihood-based procedures for
comparing rival specifications. Later, Davidson and MacKinnon \cite{davidson1981} proposed the
classical \(J\) test for non-nested regression models, exploiting the
fact that the fitted linear predictor of one model can be incorporated
as an additional regressor in the competing model. Vuong\cite{vuong1989}
subsequently developed a likelihood-ratio-based framework that unified
non-nested hypothesis testing under a general asymptotic theory
and proposed a signed, $N(0,1)$-distributed statistic
for model selection between two candidates. More recently, Hagemann
\cite{hagemann2012} introduced the minimum \(J\)
(\(MJ\)) test, extending
the classical \(J\) test to settings involving \(M\geq 2\)
competing models and avoiding the need for sequential pairwise testing.
In a broader modeling context, Cribari-Neto and Lucena \cite{cribari2017} generalized both the
\(J\) and \(MJ\) tests to generalized additive models for location, scale,
and shape (GAMLSS), where each distributional parameter is associated
with its own regression submodel. Their approach augments each submodel
with the fitted linear predictor from the corresponding competing
specification and tests the exclusion of these additional terms.

While elegant and effective in regression settings, existing \(J\)-type
methodologies are not directly applicable to an important class of
problems in survival analysis. The approach of Cribari-Neto and Lucena
\cite{cribari2017} requires at least one distributional parameter to be
modeled through a regression submodel and cannot accommodate settings in
which all parameters are fixed scalars. Moreover, it cannot be
implemented when the competing models differ exclusively in their assumed
baseline distribution, since the fitted linear
predictor of one model then reduces to a linear combination of
covariates already present in the other, rendering the augmented model
collinear. This restriction is especially limiting in cure-fraction
survival analysis, where the main inferential challenge often lies
precisely in selecting the most appropriate distribution for the
susceptible subpopulation, and where a regression
structure for the cure probability or the scale parameters may not
always be the object of study.

These limitations call for a different strategy for
constructing the augmented model. Rather than adding the fitted linear
predictor of the competing specification as an extra regressor, we
propose to augment the null log-likelihood itself with the individual
log-likelihood differences between the competing models, evaluated at
their respective maximum likelihood estimates. The redundancy of this
additional information is then assessed using a score test. The score
test is constructed solely from the restricted maximum likelihood
estimates under the null hypothesis---that is, under the assumption
that a given candidate distribution is the correct data-generating
model---without requiring estimation of the augmented
model. This is a critical advantage: because the augmented
log-likelihood in this setting is not a genuine parametric model, the
likelihood-ratio and Wald statistics do not have their standard
$\chi^2$ asymptotic distribution, while the score statistic, evaluated
at the restriction, remains valid.
Furthermore, because the score statistic reduces to a
quadratic form in the sample mean of the individual log-likelihood
differences, its signed square root coincides formally with the
statistic proposed in Vuong \cite{vuong1989}. The key departures from Vuong's approach
are threefold: we test the specific null hypothesis that a given model
is correctly specified, rather than the hypothesis of model equivalence;
we use the unsigned, $\chi^2_1$-distributed version of the statistic,
which extends naturally to \(M\geq 2\) candidates; and we estimate the
Kullback--Leibler (KL) bias present in the statistic via parametric
bootstrap, a correction that is necessary to control the type~I error
rate and that Vuong's procedure does not perform.

The main contribution of this paper is to develop a new score-based
\(J\)/\(MJ\)-type testing framework for non-nested parametric survival
models, with particular emphasis on cure-fraction settings. The central
idea is to augment the null log-likelihood score
equation with individual log-likelihood differences evaluated at the
maximum likelihood estimates (MLEs) of the competing
models, rather than with fitted linear predictors as in the classical
\(J\) test. This construction yields a unified and flexible testing
framework that applies directly to non-nested parametric survival
distributions, such as Weibull, log-logistic, and Gompertz models, and
remains valid when the candidate specifications differ solely in the
baseline distribution of the susceptible population.
A single set of parametric bootstrap replicates,
generated under the model assumed correct by each null hypothesis,
simultaneously provides a consistent estimate of the KL bias and
calibrates the critical values, so that the correction incurs no
additional simulation cost. To the best of our knowledge, this is the
first extension of \(J\)- and \(MJ\)-type tests to this class of survival
models.

The remainder of the paper is organized as follows. Section~2 develops the proposed methodology. Section~2.1 introduces the general score-test framework, while Section~2.2 discusses its specific advantages for non-nested hypothesis testing. Section~2.3 presents the proposed construction for models that differ only in their baseline distribution, which constitutes the primary setting of interest in this paper. The connection between the proposed methodology and Vuong's test is examined in Section~2.4, Section~2.5 extends the framework to \(M\geq 2\) competing models, and Section~2.6 adapts the approach to survival models with a cure fraction. Section~3 establishes the asymptotic theory, whereas Section~4 describes bootstrap implementations of the individual \(J\) tests and the global \(MJ\) test. Monte Carlo evidence is reported in Section~5, and an empirical application to cervical cancer survival data is presented in Section~6. Concluding remarks are offered in Section~7.

\section{Hypothesis Tests for Non-Nested Survival Models}
\label{sec:methodology}

\subsection{The score test}
\label{subsec:score_formula}

Consider a parametric model with log-likelihood $\ell(\btheta,\bgamma)$,
where $\btheta\in\mathbb{R}^p$ collects the parameters of interest and
$\bgamma\in\mathbb{R}^q$ are additional parameters whose exclusion is to
be tested. The null hypothesis is $H_0:\bgamma=\mathbf{0}$. Let
$\tilde{\btheta} = \argmax_{\btheta}\,\ell(\btheta,\mathbf{0})$ denote
the restricted MLE, obtained by maximizing the log-likelihood subject to
$H_0$. The score vector of the additional parameters, evaluated at the
restricted estimator, is
\begin{equation*}
  \mathbf{S}_{\bgamma}
  = \frac{\partial\ell}{\partial\bgamma}
    \bigg|_{\btheta=\tilde{\btheta},\,\bgamma=\mathbf{0}}
  = \sum_{i=1}^n
    \frac{\partial\ell_i}{\partial\bgamma}
    \bigg|_{\tilde{\btheta},\,\mathbf{0}}
  \in \mathbb{R}^q,
\end{equation*}
where $\ell_i$ is the $i$th individual contribution to the
log-likelihood. The score statistic for $H_0:\bgamma=\mathbf{0}$ is
\begin{equation}
  S = \mathbf{S}_{\bgamma}^\top\,
      \mathbf{K}_{\bgamma\bgamma\cdot\btheta}^{-1}\,
      \mathbf{S}_{\bgamma},
  \label{eq:score_stat}
\end{equation}
where $\mathbf{K}_{\bgamma\bgamma\cdot\btheta}$ is the partial
Fisher information matrix of $\bgamma$, obtained by the block-matrix
inversion formula applied to the full Fisher information matrix
$\mathbf{K}$ after eliminating the contribution of $\btheta$:
\begin{equation*}
  \mathbf{K}_{\bgamma\bgamma\cdot\btheta}
  = \mathbf{K}_{\bgamma\bgamma}
  - \mathbf{K}_{\bgamma\btheta}\,\mathbf{K}_{\btheta\btheta}^{-1}\,
    \mathbf{K}_{\btheta\bgamma}.
\end{equation*}
Here $\mathbf{K}$ is the $(p+q)\times(p+q)$ Fisher information matrix,
partitioned conformably with $(\btheta^\top,\bgamma^\top)^\top$. Since
$\mathbf{K}_{\bgamma\bgamma\cdot\btheta}$ is positive semi-definite and
$S$ is a quadratic form, $S\geq 0$. Under $H_0$ and standard regularity
conditions, $S\xrightarrow{d}\chi^2_q$ as $n\to\infty$, where
$q=\dim(\bgamma)$.

For right-censored survival data, the expected Fisher information
$\mathbf{K} = \E[-\nabla^2_{\btheta,\bgamma}\,\ell]$ rarely has a
closed form, as it involves integration over the censoring
distribution. In this case $\mathbf{K}$ is replaced by the observed
information matrix,
\begin{equation*}
  \hat{\mathbf{K}}
  = -\nabla^2_{\btheta,\bgamma}\,\ell(\btheta,\bgamma)
    \big|_{\tilde{\btheta},\,\mathbf{0}},
\end{equation*}
which satisfies $n^{-1}(\hat{\mathbf{K}} - \mathbf{K})\xrightarrow{p}
\mathbf{0}$ under standard conditions, so that the asymptotic
$\chi^2_q$ distribution of $S$ under $H_0$ is preserved.

\subsection{Advantages of the score test in the non-nested context}
\label{subsec:score_vantagens}

The score statistic~\eqref{eq:score_stat} is computed entirely from the
restricted model, evaluated at the restricted MLE $\tilde{\btheta}$.
As a result, fitting the augmented model is not required to obtain the
test statistic: the competing models are needed only to construct the
additional variables, and only their fitted values enter the calculation,
not a new optimization.

The likelihood-ratio and Wald statistics, by contrast, require fitting
the full augmented model. In the non-nested context of
Section~\ref{subsec:cenario_B} below, the augmented model is not a
genuine parametric model, it depends on estimated
quantities from the competing model fitted to the same data, so the
$\chi^2$ limiting distribution of the likelihood-ratio statistic is not
guaranteed by Wilks's theorem. The score test circumvents this
difficulty because the statistic is evaluated at the restricted
estimator, where the log-likelihood reduces to that of the genuine
restricted model.

Further advantages of the score test include invariance to
reparametrizations of $\bgamma$, superior finite-sample performance
when the augmented model is ill-conditioned, and applicability when the
augmented model fails to converge numerically.

\subsection{Scenario: models differing only in the baseline
distribution}\label{subsec:cenario_B}

The \(J\) test, introduced by Davidson and MacKinnon \cite{davidson1981}   for non-nested
regression models, operates as follows. Suppose two candidate models,
$\mathcal{M}_1$ and $\mathcal{M}_2$, are indexed by linear predictors
that involve different sets of covariates or functionally distinct
specifications. Under $\mathcal{M}_1$, the fitted linear predictor of
$\mathcal{M}_2$ provides genuine additional information not captured by
$\mathcal{M}_1$. The \(J\) test therefore augments $\mathcal{M}_1$ by
introducing the fitted linear predictor
$\hat\varphi_i^{(2)} = \mathbf{X}_{2i}^\top\hat{\bbeta}_2$
from $\mathcal{M}_2$ as an extra covariate, where
$\mathbf{X}_{2i}\in\mathbb{R}^{p_2}$ is the covariate vector of the
$i$th observation under $\mathcal{M}_2$ and
$\hat{\bbeta}_2\in\mathbb{R}^{p_2}$ is the corresponding MLE. A test
on the coefficient of $\hat\varphi_i^{(2)}$ then assesses whether
$\mathcal{M}_2$ contributes significantly beyond $\mathcal{M}_1$.

In survival analysis, however, a fundamental first step is to identify
the probability distribution that best fits the survival time itself,
before any regression structure is introduced. The choice of baseline
distribution---whether Weibull, log-normal, Gompertz, or another
parametric form---directly affects subsequent inference, including the
specification of regression models and, in cure-fraction settings, the
estimation of cure probabilities. Thus, our primary interest lies in
comparing non-nested probabilistic models for the survival time.

When no regression structure is present, there are no
linear predictors to augment and the classical \(J\) test is not directly
applicable. 
We therefore propose a different augmentation
strategy.

\subsubsection{The stimulated log-likelihood approach}

The solution we propose is to replace $\hat\varphi_i^{(2)}$ by
\begin{equation}
  q_i = \hat\ell_{i2} - \hat\ell_{i1},\quad i=1,\ldots,n,
  \label{eq:qi}
\end{equation}
where $\hat\ell_{im} = \ell_{im}(\htheta_m)$ is the individual
log-likelihood of model $\mathcal{M}_m$ evaluated at its MLE. The
quantity $q_i$ measures, for the $i$th observation, how much better
$\mathcal{M}_2$ fits relative to $\mathcal{M}_1$, using the full
distributional information rather than only the linear predictor. It
therefore captures the relevant contrast between the two candidate
baseline distributions without requiring a regression structure.

The augmented model of $\mathcal{M}_1$ is defined by the stimulated
log-likelihood 
\begin{equation}
  \ell_1^*(\btheta_1,\gamma)
  = \sum_{i=1}^n
    \bigl[\ell_{i1}(\btheta_1) + \gamma\,q_i\bigr],
  \label{eq:stimulated}
\end{equation}
where $\gamma\in\mathbb{R}$ is an additional scalar parameter. Setting
$\gamma=0$ recovers $\ell_1(\btheta_1)$; nonzero values of $\gamma$
capture the additional information about the adequacy of $\mathcal{M}_1$
contained in the observed log-likelihood differences.

This formulation differs fundamentally from the approach of Cribari-Neto
and Lucena
\cite{cribari2017}, who augment the regression submodels of
\(M_1\) with the fitted linear predictors of \(M_2\).
Our augmentation acts at the level of the log-likelihood itself, rather
than at the level of any particular submodel, so it does not require a
regression structure and remains valid when the models differ only in
their baseline distribution.

\begin{remark}
The function $\ell_1^*(\btheta_1,\gamma)$ in~\eqref{eq:stimulated} is
not a genuine log-likelihood, since $q_i$ depends on $\htheta_2$,
which is estimated from the same data. Accordingly, $\gamma$ is not a
parameter in the strict sense, and the $\chi^2_1$ limit of the
likelihood-ratio statistic $2(\hat\ell_1^* - \hat\ell_1)$, with
$\ell_1(\btheta_1)= \sum_{i=1}^n\ell_{i1}(\btheta_1)$, is not
guaranteed by Wilks's theorem. The score test avoids this difficulty:
the statistic is evaluated at $\gamma=0$, where
$\ell_1^*(\btheta_1,0) = \ell_1(\btheta_1)$ is a genuine
log-likelihood and the asymptotic theory applies.
\end{remark}

\subsubsection{The score statistic}

The score of $\ell_1^*$ with respect to $\gamma$, evaluated at the
restricted MLE $(\htheta_1^\top,0)^\top$, is
\begin{equation*}
  S_\gamma^{(1)}
  = \frac{\partial\ell_1^*}{\partial\gamma}
    \bigg|_{\htheta_1,\,0}
  = \sum_{i=1}^n q_i = n\bar{q},
\end{equation*}
where $\bar{q} = n^{-1}\sum_i q_i$. To construct the score test
statistic, we need the partial information of $\gamma$ given $\btheta_1$.
The information matrix of $\ell_1^*$ in the full parameter vector
$(\btheta_1^\top,\gamma)^\top$, evaluated at $(\htheta_1^\top,0)^\top$,
takes the block form
\begin{equation*}
  \hat{\mathbf{K}} =
  \begin{pmatrix}
    \hat{\mathbf{K}}_{\btheta_1\btheta_1}
      & \hat{\mathbf{k}}_{\btheta_1\gamma} \\[4pt]
    \hat{\mathbf{k}}_{\btheta_1\gamma}^\top
      & \hat{K}_{\gamma\gamma}
  \end{pmatrix},
\end{equation*}
with $\hat{\mathbf{K}}_{\btheta_1\btheta_1}\in\mathbb{R}^{p_1\times p_1}$,
$\hat{\mathbf{k}}_{\btheta_1\gamma}\in\mathbb{R}^{p_1}$, and
$\hat{K}_{\gamma\gamma}\in\mathbb{R}$. Direct differentiation gives
\begin{align*}
  \hat{K}_{\gamma\gamma}
    &= -\frac{\partial^2\ell_1^*}{\partial\gamma^2}
       \bigg|_{\htheta_1,\,0}
     = \sum_{i=1}^n q_i^2
     \approx n\,\widehat{\Var}(q_i),
  \\[4pt]
  \hat{\mathbf{k}}_{\btheta_1\gamma}
    &= -\frac{\partial^2\ell_1^*}
             {\partial\btheta_1\,\partial\gamma}
       \bigg|_{\htheta_1,\,0}
     = -\sum_{i=1}^n q_i\,\mathbf{s}_{i1},
\end{align*}
where $\mathbf{s}_{i1} = \partial\ell_{i1}/\partial\btheta_1
|_{\htheta_1}\in\mathbb{R}^{p_1}$ is the individual score vector of
$\mathcal{M}_1$. The partial information of $\gamma$ given $\btheta_1$
is then
\begin{equation}
  \hat{K}_{\gamma\gamma\cdot\btheta_1}
  = n\,\widehat{\Var}(q_i)
  - n\,\widehat{\Cov}(q_i,\mathbf{s}_{i1})^\top
    \hat{\mathbf{K}}_{\btheta_1\btheta_1}^{-1}
    \widehat{\Cov}(q_i,\mathbf{s}_{i1}),
  \label{eq:Kgg_parcial_exp}
\end{equation}
where $\widehat{\Cov}(q_i,\mathbf{s}_{i1}) =
n^{-1}\sum_i(q_i-\bar{q})(\mathbf{s}_{i1}-\bar{\mathbf{s}}_1)
\in\mathbb{R}^{p_1}$. The score statistic is
\begin{equation*}
  J_1^{\mathrm{sand}}
  = \frac{(S_\gamma^{(1)})^2}{\hat{K}_{\gamma\gamma\cdot\btheta_1}}
  = \frac{n^2\bar{q}^2}{\hat{K}_{\gamma\gamma\cdot\btheta_1}}
  \xrightarrow{d} \chi^2_1
  \quad\text{under }H_0^{(1)},
\end{equation*}
where the limit follows from the theory of $M$-estimators
\cite{white1982,white1994}.

\subsubsection{An alternative information estimator}
\label{subsubsec:info_alt}

The correction term in~\eqref{eq:Kgg_parcial_exp} can be numerically
unstable in moderate samples. When $\mathcal{M}_1$ is correctly
specified, the quantities $q_i$ and the scores $\mathbf{s}_{i1}$ tend
to be highly correlated, driving the quadratic form
$n\,\widehat{\Cov}(q_i,\mathbf{s}_{i1})^\top
\hat{\mathbf{K}}_{\btheta_1\btheta_1}^{-1}
\widehat{\Cov}(q_i,\mathbf{s}_{i1})$ close to $n\,\widehat{\Var}(q_i)$
and the denominator of $J_1^{\mathrm{sand}}$ toward zero. Monte Carlo
experiments confirm that the resulting test statistic exhibits severely
inflated kurtosis and the test has poor size control in finite samples.

A numerically stable alternative is to omit the correction term and
use
\begin{equation*}
  \hat{K}_{\gamma\gamma\cdot\btheta_1}^{\mathrm{alt}}
  = n\,\widehat{\Var}(q_i)
  = \frac{n}{n-1}\sum_{i=1}^n(q_i-\bar{q})^2
  \equiv n\,\hat\sigma_q^2.
\end{equation*}
The resulting test statistic is
\begin{equation}
  J_1 = \frac{n\bar{q}^2}{\hat\sigma_q^2},
  \label{eq:J1_proposta}
\end{equation}
which retains the $\chi^2_1$ asymptotic null distribution because
$\hat\sigma_q^2$ is a consistent estimator of
$\sigma_W^2 = \Var_{\btheta_1^0}(q_i)$, the asymptotic variance of
$\sqrt{n}\bar{q}$ under $H_0^{(1)}$ (Proposition~\ref{prop:asym}).
Monte Carlo evidence confirms that~\eqref{eq:J1_proposta} achieves
substantially better size control in finite samples than
$J_1^{\mathrm{sand}}$.

\subsection{Relationship with the Vuong test and contributions of the
proposed approach}
\label{subsec:vuong}

The test statistic~\eqref{eq:J1_proposta} bears a
direct formal relationship to the likelihood-ratio-based test proposed
by Vuong~\cite{vuong1989}. Vuong's statistic for
discriminating between two non-nested models \textcolor{blue}{is}
\begin{equation*}
  V_n = \frac{\sqrt{n}\,\bar{q}}{\hat\sigma_q}
  \xrightarrow{d} N(0,1)
  \quad\text{under }H_0^V,
\end{equation*}
where \(H^V_0: \mathbb{E}[q_i]=0\), i.e., the two models fit the data
equally well in Kullback-Leibler (KL) distance to the true distribution
\cite{vuong1989,kullback1951}. Positive values of $V_n$ favor $\mathcal{M}_2$;
negative values favor $\mathcal{M}_1$.

Since $J_1 = V_n^2$, statistic~\eqref{eq:J1_proposta} is the square of
the Vuong statistic. Despite this formal connection, the proposed
approach differs from Vuong's in three fundamental respects.

First, the null hypotheses are different. Vuong tests whether the two
models are equidistant from the true distribution in KL terms. This null hypothesis
can hold when both models are misspecified with equal KL divergences, and
it is necessarily false when exactly one model is the true
data-generating process, since that model's KL divergence is zero while
the other's is strictly positive. Consequently, the Vuong test cannot
assess whether a specific model is correctly specified, and cannot signal
that all candidates are misspecified. We instead test $H_0^{(m)}$:
$\mathcal{M}_m$ is the true data-generating process, a hypothesis with
a direct substantive interpretation.

Second, the Vuong test has distorted size when used as a test of
$H_0^{(m)}$. Under $H_0^{(1)}$, Proposition~\ref{prop:asym} shows that
$\bar{q}\xrightarrow{p}\mu_W = -\KL(\mathcal{M}_1\|\mathcal{M}_2)<0$,
so $V_n\to-\infty$ and the rejection rate of $H_0^V$ collapses to zero
precisely when $\mathcal{M}_1$ is the true model. By estimating $\mu_W$
via parametric bootstrap and centering the statistic accordingly, we
restore the correct $\chi^2_1$ null distribution.

Third, the Vuong test is confined to two models. Because $V_n$ is
signed, it has no natural analogue for $M\geq 3$ candidates.
Statistic~\eqref{eq:J1_proposta} is unsigned and therefore coherent for
any \(M\geq 2\): \(J_m\) measures how far $\bar{q}^{(m)}$ deviates from
its null expectation regardless of direction.

\subsection{\texorpdfstring{Extension to \(M\geq 2\) candidate models}{Extension to M >= 2 candidate models}}

\label{subsec:extension}

The preceding development extends straightforwardly to \(M\geq 2\)
candidate models. For each $\mathcal{M}_m$, $m=1,\ldots,M$, define
the individual log-likelihood differences relative to the best
competing model:
\begin{equation}
  q_i^{(m)} = \max_{m'\neq m}\hat\ell_{im'} - \hat\ell_{im},
  \quad i=1,\ldots,n.
  \label{eq:qi_M}
\end{equation}
Set $\bar{q}^{(m)} = n^{-1}\sum_i q_i^{(m)}$ and
$\hat\sigma_{q^{(m)}}^2 = (n-1)^{-1}\sum_i(q_i^{(m)}-\bar{q}^{(m)})^2$.
The stimulated log-likelihood of $\mathcal{M}_m$ is
$\ell_m^*(\btheta_m,\gamma) = \sum_i[\ell_{im}(\btheta_m) +
\gamma\,q_i^{(m)}]$, and the score at $(\htheta_m^\top,0)^\top$ is
$S_\gamma^{(m)} = n\bar{q}^{(m)}$. Using the alternative information
estimator $n\hat\sigma_{q^{(m)}}^2$, the \(J_m\) statistic is
\begin{equation*}
  J_m = \frac{n(\bar{q}^{(m)}-\hat\mu_m)^2}{\hat\sigma_{q^{(m)}}^2},
  \quad m=1,\ldots,M,
\end{equation*}
where $\hat\mu_m$ is the bootstrap estimate of the KL bias
$\mu_m = \E_{\btheta_m^0}[\ell_{im'}(\bphi_{m'}^*)] -
\E_{\btheta_m^0}[\ell_{im}(\btheta_m^0)]$, described in
Section~\ref{subsec:mu_est}. Each $J_m\xrightarrow{d}\chi^2_1$ under
$H_0^{(m)}$.

When \(M\geq 2\) candidate models are under
consideration,Hagemann~\cite{hagemann2012} proposed combining the individual
\(J_m\) statistics into a single global test to avoid sequential pairwise
testing. The \(MJ\) statistic is defined as
\begin{equation}
  MJ = \min_{m=1,\ldots,M} J_m.
  \label{eq:MJ}
\end{equation}
Since each $J_m\geq 0$, the minimum requires no absolute values, and
the definition is coherent for any $M$. The global null hypothesis is
$H_0$: there exists $m^*\in\{1,\ldots,M\}$ such that $\mathcal{M}_{m^*}$
is the true data-generating process. The test outcome is interpreted as
follows:
\begin{itemize}
  
\item If \(p_{MJ}>\alpha\), \(H_0\) is not rejected and the model
\(\widehat{m}=\argmin_{m=1,\ldots,M} J_m\) is selected, that is, the
candidate whose null hypothesis is least contradicted by the data.
  \item If $p_{MJ}\leq\alpha$, $H_0$ is rejected, indicating that the
        true data-generating process is not among the $M$ candidates and
        that the model set should be expanded.
\end{itemize}

\begin{remark}
The model selected by \(\argmin_m J_m\) and the model selected by
\(\argmax_m \widehat{\ell}_m\) need not coincide in finite samples, since
\(J_m\) depends not only on the maximized log-likelihoods, but also on the
bootstrap estimate \(\widehat{\mu}_m\) of the KL bias and on
\(\widehat{\sigma}_{q^{(m)}}^2\). In large samples, however, the two criteria
are expected to agree with probability approaching one when one of the candidate
models is correctly specified, because the correctly specified model tends to
have both the largest maximized log-likelihood and the smallest centered
log-likelihood contrast.
\end{remark}

Note that, unlike AIC and BIC, the \(MJ\) test is the
only procedure in this framework capable of indicating that all
candidate models are inadequate. If all individual tests \(J_m\) reject
their respective null hypotheses, there is evidence that every candidate
model is misspecified, and the analyst should consider a different class of parametric distributions.

Table~\ref{tab:comparison} summarizes the key differences between the
proposed approach and the Vuong test.

\begin{table}[!ht]
\footnotesize
\centering
\caption{Comparison of the proposed approach and Vuong's test.}
\label{tab:comparison}
\begin{tabular}{p{2.1cm} p{2.8cm} p{2.8cm}}
\toprule
\textbf{Feature} & \textbf{Vuong (1989)} & \textbf{Proposed-approach} \\
\midrule
Null hypothesis & Equivalence ($\mu_W=0$) & $\mathcal{M}_m$ is the true model \\
Test statistic & Signed, $N(0,1)$ & Unsigned, $\chi^2_1$ \\
Number of models & $M=2$ & \(M\geq 2\) \\
KL bias correction & No & Yes (bootstrap) \\
Reject all models & No & Yes (\(MJ\) test) \\
Model selection & Partial (sign of \(V_n\)) & Yes (\(\argmin_m J_m\)) \\
Regression required & No & No \\
\bottomrule
\end{tabular}
\end{table}

\subsection{Survival models with a cure fraction}
\label{subsec:cure_models}

Long-term survival models, also known as cure-fraction models, are
commonly formulated through the standard mixture representation
introduced in the cancer survival literature
~\cite{boag1949,berk,maller1996,tso,rondeau2013, su2022}. In this formulation, the
population is viewed as a mixture of cured individuals and susceptible
individuals, so that the population survival function converges to a
positive limit as $t\to\infty$. 

To illustrate the proposed framework,
consider $M=2$ candidate models: a Weibull long-term (Weibull-LT) model
and a Gompertz long-term (Gompertz-LT) model. Under both, the
population survival function is
\begin{equation*}
  S_m(t;\btheta_m) = p_{0m} + (1-p_{0m})\,S_m^*(t;\balpha_m),
\end{equation*}
where $p_{0m}\in(0,1)$ is the cure fraction and $S_m^*$ is the
survival function of the susceptible subpopulation. The individual
log-likelihood contribution is
\begin{equation*}
  \ell_{im}(\btheta_m) =
  \begin{cases}
    \log(1-p_{0m}) + \log f_m^*(T_i;\balpha_m), & \delta_i=1,\\[4pt]
    \log\bigl[p_{0m}+(1-p_{0m})\,S_m^*(T_i;\balpha_m)\bigr],
    & \delta_i=0.
  \end{cases}
\end{equation*}

Under the Weibull-LT model, \(S_1^*(t)=\exp(-\lambda t^k)\), with
\(k,\lambda>0\), where the susceptible component follows a Weibull
distribution ~\cite{weibull1951}, and
\(\btheta_1=(k,\lambda,p_{01})^\top\). Under the Gompertz-LT model~\cite{gieser}, \(S_2^*(t)=
\exp\{-(\alpha/\xi)(\mathrm{e}^{\xi t}-1)\}\)
with \(\alpha,\xi>0\),
and \(\btheta_2=(\alpha,\xi,p_{02})^\top\). Both
models are indexed by three parameters and differ only in the baseline
distribution of the susceptible subpopulation. Therefore, the
construction developed in Section~\ref{subsec:cenario_B} applies
directly, and the $q_i$ in~\eqref{eq:qi} are defined through the
Weibull-LT and Gompertz-LT individual log-likelihoods. The
extension to a third candidate model, such as the Log-logistic-LT, is
immediate: one replaces~\eqref{eq:qi} by~\eqref{eq:qi_M} and applies
the $M=3$ version of the \(MJ\) statistic~\eqref{eq:MJ}.

\section{Asymptotic Distribution Theory}
\label{sec:asym}

We state the main asymptotic result for the case $M=2$; the extension
to $M>2$ is analogous. Let $\btheta_1^0$ denote the true parameter value
under $H_0^{(1)}$, and let $\bphi^*$ be the pseudo-true value of
$\mathcal{M}_2$:
$\bphi^* = \argmax_{\bphi}\,\E_{\btheta_1^0}[\ell_{i2}(\bphi)]$.
The vector $\bphi^*$ is the probability limit of the MLE of
$\mathcal{M}_2$ when the true distribution is $\mathcal{M}_1$, and it
minimizes $\KL(\mathcal{M}_1\|\mathcal{M}_2)$ over the parametric family
of $\mathcal{M}_2$ \cite{white1982}.

\begin{proposition}
\label{prop:asym}
Under standard regularity conditions, if $\mathcal{M}_1$ is the true
data-generating process:
\begin{enumerate}[label=(\roman*)]
  \item $\bar{q}\xrightarrow{p}
        \mu_W
        = \E_{\btheta_1^0}[\ell_{i2}(\bphi^*)]
          - \E_{\btheta_1^0}[\ell_{i1}(\btheta_1^0)]
        = -\KL(\mathcal{M}_1\|\mathcal{M}_2)\leq 0$;
  \item $\sqrt{n}(\bar{q}-\mu_W)\xrightarrow{d}N(0,\sigma_W^2)$
        with $\sigma_W^2=\Var_{\btheta_1^0}(q_i)$;
  \item $J_1 = n(\bar{q}-\hat\mu_W)^2/\hat\sigma_q^2
        \xrightarrow{d}\chi^2_1$ under $H_0^{(1)}$, where $\hat\mu_W$
        is the bootstrap estimator of $\mu_W$ defined in
        Section~\ref{subsec:mu_est}.
\end{enumerate}
\end{proposition}

Item~(i) holds because $\mathcal{M}_2$, fitted with pseudo-true
parameters $\bphi^*$ rather than the true $\btheta_1^0$, cannot exceed
the expected log-likelihood of the correctly specified $\mathcal{M}_1$.
The inequality $\mu_W\leq 0$ is an equality only if
$\mathcal{M}_2$ is also correctly specified, which is impossible for
genuinely non-nested models; hence $\mu_W<0$ strictly.
Therefore, the uncentered Vuong statistic
$V_n = \sqrt{n}\bar{q}/\hat\sigma_q\to-\infty$ under $H_0^{(1)}$,
confirming that the Vuong test does not control the type~I error rate
when used to assess whether $\mathcal{M}_1$ is the true model.

\subsection{Estimation of the KL bias}
\label{subsec:mu_est}

Since $\mu_W$ is unknown, it is estimated by parametric bootstrap under
$\mathcal{M}_1(\htheta_1)$. Specifically, generate $B$
independent samples
$\mathbf{Z}^{(1)},\ldots,\mathbf{Z}^{(B)}$ from $\mathcal{M}_1$ with
parameters $\htheta_1$. Fit both models to each sample, compute
the corresponding log-likelihood differences $q_i^{(b)}$ and their mean
$\bar{q}^{(b)}$, and set
\begin{equation*}
  \hat\mu_W = \frac{1}{B}\sum_{b=1}^B \bar{q}_W^{(b)}.
\end{equation*}
As $B$ and $n$ jointly diverge, $\hat\mu_W$
converges in probability to $\mu_W$, and the joint consistency of
$(\hat\mu_W,\hat\sigma_q^2)$ implies item~(iii) of
Proposition~\ref{prop:asym}.

Analogously, $\mu_G$ is estimated by generating $B$
independent samples from $\mathcal{M}_2(\htheta_2)$, fitting both
models to each sample, and computing
\begin{equation*}
  \hat\mu_G = \frac{1}{B}\sum_{b=1}^B \bigl(-\bar{q}_G^{(b)}\bigr),
\end{equation*}
where $\bar{q}_G^{(b)} = n^{-1}\sum_i(\hat\ell_{i2}^{(b)} -
\hat\ell_{i1}^{(b)})$ is the mean log-likelihood difference in the
$b$th sample generated under $\mathcal{M}_2$.

The proposed test statistics for $H_0^{(1)}$ and $H_0^{(2)}$ are
\begin{equation*}
  J_1 = \frac{n(\bar{q}-\hat\mu_W)^2}{\hat\sigma_q^2}
  \,\, \text{ and } 
  J_2 = \frac{n(-\bar{q}-\hat\mu_G)^2}{\hat\sigma_q^2},
\end{equation*}
where \(-\bar q\) in \(J_2\) reflects the reversal of the comparison direction
when \(M_2\) is the null model. Both
statistics share the same denominator because $\Var(-q_i)=\Var(q_i)$,
but have distinct numerators since $\hat\mu_W\neq\hat\mu_G$ in general:
the two KL divergences, $\KL(\mathcal{M}_1\|\mathcal{M}_2)$ and
$\KL(\mathcal{M}_2\|\mathcal{M}_1)$, are generally different.
Asymptotic $p$-values are $p_m^{\mathrm{asymp}}=1-F_{\chi^2_1}(J_m)$.

A key practical feature of the procedure is that the
same $B$ bootstrap replicates generated under $\mathcal{M}_m$ for
bias estimation can simultaneously be used to calibrate the critical
values of \(J_m\), as described in Section~\ref{subsec:boot_J}. No
additional simulation is therefore required to obtain bootstrap
$p$-values.

\section{Bootstrap Calibration}
\label{sec:bootstrap}

In finite samples, particularly with heavy censoring
or extreme cure fractions, the $\chi^2_1$ approximation to the null
distribution of \(J_m\) may be inadequate. Parametric bootstrap
calibration avoids reliance on this approximation
\cite{Davison1997} and, crucially, uses the same $B$ replicates
already generated in Section~\ref{subsec:mu_est} for KL-bias
estimation at no additional computational cost. We describe the
bootstrap procedures for the individual \(J_m\) tests and for the global
\(MJ\) test in turn.

\subsection{\texorpdfstring{Individual \(J\) tests}{Individual J tests}}
\label{subsec:boot_J}

\paragraph{Procedure for \(J_1\).}
\begin{enumerate}[label=\arabic*.]
  \item Compute $\htheta_1$, $\htheta_2$, $q_i$, $\bar{q}$, and
        $\hat\sigma_q$ from the original sample.
  \item For $b=1,\ldots,B$, generate $\mathbf{Z}^{(b)}$ from
        $\mathcal{M}_1(\htheta_1)$ with estimated
        censoring rate
        $\hat\lambda_c = \sum_i(1-\delta_i)/\sum_i T_i$; fit both
        models; obtain $\bar{q}^{(b)}$ and $\hat\sigma_q^{(b)}$.
  \item Compute $\hat\mu_W = B^{-1}\sum_b\bar{q}^{(b)}$ and the
        bootstrap statistics
        $J_1^{(b)} = n(\bar{q}^{(b)}-\hat\mu_W)^2/
        (\hat\sigma_q^{(b)})^2$.
  \item The bootstrap $p$-value is
        \begin{equation*}
          p_1^{\mathrm{boot}}
          = \frac{1+\#\{b:J_1^{(b)}\geq J_1\}}{B+1},
        \end{equation*}
        where $J_1 = n(\bar{q}-\hat\mu_W)^2/\hat\sigma_q^2$.
        Reject $H_0^{(1)}$ if $p_1^{\mathrm{boot}}\leq\alpha$.
\end{enumerate}

The procedure for $J_2$ is analogous, with bootstrap samples generated
under $\mathcal{M}_2(\htheta_2)$ and
$J_2^{(b)} = n(-\bar{q}^{(b)}-\hat\mu_G)^2/(\hat\sigma_q^{(b)})^2$.
For $M>2$ candidate models, the procedure generalizes
directly: for each $\mathcal{M}_m$, one generates $B$ samples under
$\mathcal{M}_m(\htheta_m)$, computes $q_i^{(m,b)}$ as
in~\eqref{eq:qi_M}, estimates $\hat\mu_m$ as the mean of
$\bar{q}^{(m,b)}$ across replicates, and defines $J_m^{(b)}$
accordingly.

\subsection{\texorpdfstring{The \(MJ\) test}{The MJ test}}
\label{subsec:boot_MJ}

The \(p\)-value of the \(MJ\) test is calibrated using the procedure of
Hagemann~\cite{hagemann2012}, which we adapt here to the bootstrap statistics
defined above.

\paragraph{Procedure.}
\begin{enumerate}[label=\arabic*.]
  \item Compute $\hat{m} = \argmin_{m=1,\ldots,M}J_m$; set
        $MJ_{\mathrm{obs}} = J_{\hat{m}}$.
  \item Using the $B$ bootstrap replicates already
        generated under $\mathcal{M}_{\hat{m}}(\htheta_{\hat{m}})$ in
        step~2 of Section~\ref{subsec:boot_J}, compute, for each
        replicate $b$,
        $J_{\hat{m}}^{*(b)} = n(\bar{q}^{(b)}-\hat\mu_{\hat{m}})^2/
        (\hat\sigma_q^{(b)})^2$.
  \item Form the bootstrap \(MJ\) statistic:
        \begin{equation*}
          MJ^{*(b)}
          = \min\Bigl(J_{\hat{m}}^{*(b)},\;
            \min_{m\neq\hat{m}} J_m\Bigr),
        \end{equation*}
        where \(J_m\), $m\neq\hat{m}$, are the values from the original
        sample.
  \item The bootstrap $p$-value is
        \begin{equation*}
          p_{MJ}^{\mathrm{boot}}
          = \frac{1+\#\{b:MJ^{*(b)}\geq MJ_{\mathrm{obs}}\}}{B+1}.
        \end{equation*}
        Reject $H_0$ if $p_{MJ}^{\mathrm{boot}}\leq\alpha$.
\end{enumerate}

Step~3 restricts the bootstrap \(MJ\) statistic by the minimum of the
observed \(J_m\) values from the competing models \cite{hagemann2012}. 
This restriction improves finite-sample
size and power by capping large realizations of
$J_{\hat{m}}^{*(b)}$. Intuitively, when the true
model $\mathcal{M}_{\hat{m}}$ is correctly specified, the competing
statistics $J_{m'}$ ($m'\neq\hat{m}$) are expected to be large; using
their observed values as an upper bound therefore prevents the
bootstrap distribution of $MJ^{*(b)}$ from placing too much mass in
the upper tail. However, when \(\min_{m\neq \hat m} J_m\) is small
due to sampling variability, the restriction can make the test
conservative in small samples \cite{hagemann2012}. In large samples, the competing
$J_{m'}$ ($m'\neq\hat{m}$) diverge to $+\infty$ when
$\mathcal{M}_{\hat{m}}$ is the true model, so the restriction becomes
asymptotically negligible and the test is asymptotically valid.

\section{Monte Carlo Simulation}
\label{sec:simulation}

We assess the finite-sample performance of the
proposed tests through two Monte Carlo studies, examining empirical
size and power of the individual \(J_m\) tests and of the global \(MJ\)
test under both full-bootstrap (FB) and partial-bootstrap (PB)
calibration. Table~\ref{tab:parameter_values_simulation} reports the
parameter values used to generate the data. In all scenarios, the cure
fraction is fixed at \(p_{0,\mathrm{true}}=0.30\), the censoring
parameter is set to  \(\lambda_c=0.10\) and \(\lambda_c=0.30\), the number
of Monte Carlo replications is \(R_{\mathrm{MC}}=5{,}000\), and the
number of bootstrap replications is \(B=499\).

\begin{table}[h]
	\small\sf\centering
	\caption{Parameter values used in the Monte Carlo studies.}
	\label{tab:parameter_values_simulation}
	\begin{tabular}{@{}lll@{}}
		\toprule
		DGP & Study & Parameter values \\
		\midrule
		W-LT  & I, II & $k=2.60$, $\lambda=1.40$ \\
		G-LT  & I, II & $\alpha=0.20$, $\xi=1.10$ \\
		LL-LT & II    & $k=4.00$, $\lambda=0.70$ \\
		PW-LT & I, II & $h_1=0.50$, $h_2=0.001$, $h_3=6.00$,
		                $\tau_1=0.18$, $\tau_2=1.15$ \\
		\bottomrule
	\end{tabular}
\end{table}

In Study~I, the fitted candidate models are Weibull-LT and
Gompertz-LT, and samples are generated from Weibull-LT, Gompertz-LT,
and Piecewise-LT, with results reported for $n=100$, $n=250$, and
$n=500$. Study~II extends this comparison by adding Log-logistic-LT to
the fitted candidate set and also using it as an additional DGP;
because this setting involves three fitted models, results are reported
only for $n=250$. The Weibull, Gompertz, and log-logistic distributions
were chosen because they are commonly used parametric specifications for
the susceptible component in long-term survival models
\cite{boag1949,berk,maller1996}. In both studies, Piecewise-LT serves
as an external DGP, so that none of the fitted candidate
models is correctly specified when data are generated from it, allowing
assessment of the power of the \(MJ\) test to detect global
misspecification.

In Tables~\ref{tab:study_I_J_MJ_results}
and~\ref{tab:study_II_J_MJ_results}, W-LT, G-LT, LL-LT, and PW-LT
denote the Weibull-LT, Gompertz-LT, Log-logistic-LT, and Piecewise-LT
data-generating processes, respectively. The statistics
$J_{\mathrm{W}}$, $J_{\mathrm{G}}$, and $J_{\mathrm{L}}$ denote the
individual \(J_m\) tests with W-LT, G-LT, and LL-LT as the null model.
FB denotes full-bootstrap calibration, whereas PB denotes
partial-bootstrap calibration; in both cases, bootstrap resampling is
used to estimate the KL bias, but only FB also uses it to calibrate the
$p$-value. The reported quantities are empirical rejection probabilities
at the 5\% significance level.

The results in Table~\ref{tab:study_I_J_MJ_results} suggest that the
proposed tests have satisfactory finite-sample behavior in the
two-model setting. When the DGP belongs to the fitted candidate set,
the individual test associated with the correctly specified null model
has rejection probabilities close to the nominal level under FB. For
example, the rejection probabilities range from 0.047 to 0.056 for
$J_{\mathrm{W}}$ when the DGP is W-LT, and from 0.040 to 0.054 for
$J_{\mathrm{G}}$ when the DGP is G-LT. The tests associated with
misspecified null models tend to reject more often as $n$ increases.
Under FB and $\lambda_c=0.10$, for instance, the rejection probability
of $J_{\mathrm{G}}$ increases from 0.794 to 1.000 when the DGP is
W-LT, whereas that of $J_{\mathrm{W}}$ increases from 0.521 to 1.000
when the DGP is G-LT, as $n$ increases from 100 to 500. The global
test \(MJ\) also behaves in line with its intended role: it rejects
infrequently when either W-LT or G-LT is correctly specified, but
rejects more often when the DGP is PW-LT, for which neither fitted
model is correct. In this latter case, under FB, the rejection
probabilities of \(MJ\) increase from 0.498 to 0.954 for
$\lambda_c=0.10$, and from 0.573 to 0.979 for $\lambda_c=0.30$, as
$n$ increases from 100 to 500.

\begin{table}[htbp]
	\small\sf\centering
	\caption{Empirical rejection probabilities of the $J_{\mathrm{W}}$,
	$J_{\mathrm{G}}$, and \(MJ\) tests in Study~I at the 5\% significance
	level for $n=100,\,250,\,500$.}
	\label{tab:study_I_J_MJ_results}
	\begin{tabular}{@{}l|llcccc@{}}
		\toprule
		DGP & $n$ & $\lambda_c$ & Cal. & $J_{\mathrm{W}}$ & $J_{\mathrm{G}}$ & \(MJ\) \\
		\midrule
		\multirow{12}{*}{W-LT}
		& \multirow{4}{*}{100} & 0.10 & FB & 0.052 & 0.794 & 0.038 \\
		&                      &      & PB & 0.094 & 0.908 & 0.070 \\
		&                      & 0.30 & FB & 0.052 & 0.751 & 0.041 \\
		&                      &      & PB & 0.097 & 0.888 & 0.072 \\
		\cmidrule(lr){2-7}
		& \multirow{4}{*}{250} & 0.10 & FB & 0.047 & 0.994 & 0.045 \\
		&                      &      & PB & 0.072 & 0.998 & 0.070 \\
		&                      & 0.30 & FB & 0.049 & 0.986 & 0.045 \\
		&                      &      & PB & 0.084 & 0.994 & 0.080 \\
		\cmidrule(lr){2-7}
		& \multirow{4}{*}{500} & 0.10 & FB & 0.056 & 1.000 & 0.056 \\
		&                      &      & PB & 0.076 & 1.000 & 0.076 \\
		&                      & 0.30 & FB & 0.050 & 1.000 & 0.050 \\
		&                      &      & PB & 0.071 & 1.000 & 0.071 \\
		\midrule
		\multirow{12}{*}{G-LT}
		& \multirow{4}{*}{100} & 0.10 & FB & 0.521 & 0.054 & 0.004 \\
		&                      &      & PB & 0.665 & 0.102 & 0.017 \\
		&                      & 0.30 & FB & 0.411 & 0.040 & 0.001 \\
		&                      &      & PB & 0.531 & 0.085 & 0.009 \\
		\cmidrule(lr){2-7}
		& \multirow{4}{*}{250} & 0.10 & FB & 0.955 & 0.046 & 0.028 \\
		&                      &      & PB & 0.977 & 0.069 & 0.055 \\
		&                      & 0.30 & FB & 0.877 & 0.044 & 0.014 \\
		&                      &      & PB & 0.933 & 0.067 & 0.037 \\
		\cmidrule(lr){2-7}
		& \multirow{4}{*}{500} & 0.10 & FB & 1.000 & 0.047 & 0.047 \\
		&                      &      & PB & 1.000 & 0.057 & 0.057 \\
		&                      & 0.30 & FB & 0.997 & 0.042 & 0.039 \\
		&                      &      & PB & 0.999 & 0.054 & 0.053 \\
		\midrule
		\multirow{12}{*}{PW-LT}
		& \multirow{4}{*}{100} & 0.10 & FB & 0.834 & 0.498 & 0.498 \\
		&                      &      & PB & 0.873 & 0.768 & 0.767 \\
		&                      & 0.30 & FB & 0.880 & 0.573 & 0.573 \\
		&                      &      & PB & 0.912 & 0.839 & 0.839 \\
		\cmidrule(lr){2-7}
		& \multirow{4}{*}{250} & 0.10 & FB & 0.981 & 0.778 & 0.778 \\
		&                      &      & PB & 0.988 & 0.929 & 0.929 \\
		&                      & 0.30 & FB & 0.985 & 0.863 & 0.863 \\
		&                      &      & PB & 0.991 & 0.960 & 0.960 \\
		\cmidrule(lr){2-7}
		& \multirow{4}{*}{500} & 0.10 & FB & 0.997 & 0.954 & 0.954 \\
		&                      &      & PB & 0.998 & 0.974 & 0.974 \\
		&                      & 0.30 & FB & 0.999 & 0.979 & 0.979 \\
		&                      &      & PB & 0.999 & 0.992 & 0.992 \\
		\bottomrule
	\end{tabular}
\end{table}

Table~\ref{tab:study_II_J_MJ_results} shows that a similar general
pattern is observed when LL-LT is added to the candidate set. Under
FB, the tests associated with the correctly specified null model remain
close to the nominal level: the rejection probabilities are 0.029 and
0.026 for $J_{\mathrm{W}}$ when the DGP is W-LT, 0.046 and 0.047 for
$J_{\mathrm{G}}$ when the DGP is G-LT, and 0.055 and 0.044 for
$J_{\mathrm{L}}$ when the DGP is LL-LT, for $\lambda_c=0.10$ and
$\lambda_c=0.30$, respectively. The \(MJ\) test also rejects
infrequently when the fitted set contains the DGP, with FB rejection
probabilities no larger than 0.042 across W-LT, G-LT, and LL-LT.
However, when the DGP is G-LT, the rejection probabilities of
$J_{\mathrm{W}}$ are lower than in the corresponding two-model setting:
for $n=250$, these probabilities are 0.955 and 0.877 in
Table~\ref{tab:study_I_J_MJ_results}, but decrease to 0.563 and 0.547
in Table~\ref{tab:study_II_J_MJ_results} after LL-LT is added. This
suggests that including LL-LT changes the comparison used against W-LT
in the three-model setting. The diagnostic analysis in
Section~\ref{subsec:diagnostic} investigates this phenomenon.

\begin{table}[htb]
	\centering
	\caption{Empirical rejection probabilities of the $J_{\mathrm{W}}$,
	$J_{\mathrm{G}}$, $J_{\mathrm{L}}$, and \(MJ\) tests in Study~II at
	the 5\% significance level for $n=250$.}
	\label{tab:study_II_J_MJ_results}
	\begin{tabular}{@{}l|lccccc@{}}
		\toprule
		DGP & $\lambda_c$ & Cal. & $J_{\mathrm{W}}$ & $J_{\mathrm{G}}$ & $J_{\mathrm{L}}$ & \(MJ\) \\
		\midrule
		\multirow{4}{*}{W-LT}
		& 0.10 & FB & 0.029 & 0.994 & 0.996 & 0.028 \\
		&      & PB & 0.015 & 0.999 & 1.000 & 0.015 \\
		\cmidrule(lr){2-7}
		& 0.30 & FB & 0.026 & 0.989 & 0.981 & 0.026 \\
		&      & PB & 0.015 & 0.996 & 0.997 & 0.014 \\
		\midrule
		\multirow{4}{*}{G-LT}
		& 0.10 & FB & 0.563 & 0.046 & 1.000 & 0.009 \\
		&      & PB & 0.389 & 0.076 & 1.000 & 0.018 \\
		\cmidrule(lr){2-7}
		& 0.30 & FB & 0.547 & 0.047 & 1.000 & 0.003 \\
		&      & PB & 0.468 & 0.055 & 1.000 & 0.004 \\
		\midrule
		\multirow{4}{*}{LL-LT}
		& 0.10 & FB & 0.866 & 1.000 & 0.055 & 0.042 \\
		&      & PB & 0.787 & 1.000 & 0.145 & 0.077 \\
		\cmidrule(lr){2-7}
		& 0.30 & FB & 0.694 & 1.000 & 0.044 & 0.014 \\
		&      & PB & 0.625 & 1.000 & 0.096 & 0.030 \\
		\midrule
		\multirow{4}{*}{PW-LT}
		& 0.10 & FB & 0.985 & 0.755 & 0.995 & 0.759 \\
		&      & PB & 0.988 & 0.904 & 1.000 & 0.904 \\
		\cmidrule(lr){2-7}
		& 0.30 & FB & 0.990 & 0.851 & 0.999 & 0.851 \\
		&      & PB & 0.992 & 0.946 & 1.000 & 0.946 \\
		\bottomrule
	\end{tabular}
\end{table}

\subsection{Diagnostic analysis}
\label{subsec:diagnostic}

The power reduction of $J_{\mathrm{W}}$ when LL-LT is added to the
candidate set warrants closer examination, since the
mechanism is not immediately apparent from the marginal rejection rates
alone. We carried out a two-stage diagnostic analysis to understand
it.

The first stage is descriptive, based on $R_{\mathrm{MC}}=2{,}000$
Monte Carlo replications without bootstrap resampling, and examines
model selection by AIC/BIC and local features of the fitted
log-likelihood contributions. The second stage focuses on the
quantities used to construct the \(J\)-type statistics and is based on
$R_{\mathrm{MC},J}=1{,}000$ Monte Carlo replications with
$B_{\mathrm{diag}}=199$ auxiliary bootstrap samples within each
replication.

Throughout, we use the following notation. For each
observation $i$, let $\ell_{W,i}$, $\ell_{G,i}$, and $\ell_{L,i}$
denote the individual log-likelihood contributions of the fitted W-LT,
G-LT, and LL-LT models, respectively. In Study~II, define
$m_{G,L,i}=\max\{\ell_{G,i},\ell_{L,i}\}$,
$m_{W,L,i}=\max\{\ell_{W,i},\ell_{L,i}\}$,
$m_{W,G,i}=\max\{\ell_{W,i},\ell_{G,i}\}$, and
$m_i=\max\{\ell_{W,i},\ell_{G,i},\ell_{L,i}\}$, so that $m_{G,L,i}$
is the best local competitor against W-LT, $m_{W,L,i}$ is the best
local competitor against G-LT, and $m_{W,G,i}$ is the best local
competitor against LL-LT. For each null model $M$ in the fitted
candidate set, set $q_{M,i}=m_{-M,i}-\ell_{M,i}$, where $m_{-M,i}$
is the largest individual log-likelihood contribution among the
competing models; in Study~II this gives $q_{W,i}=m_{G,L,i}-
\ell_{W,i}$, $q_{G,i}=m_{W,L,i}-\ell_{G,i}$, and
$q_{L,i}=m_{W,G,i}-\ell_{L,i}$.

\paragraph{First-stage results.}
Table~\ref{tab:diag_selection} shows that AIC and BIC select the
correct model with high frequency when the DGP belongs to the
candidate set. Under the G-LT DGP in Study~II, G-LT is selected in
0.983 and 0.969 of the replications for $\lambda_c=0.10$ and
$\lambda_c=0.30$, respectively, so the power reduction in
$J_{\mathrm{W}}$ is not attributable to a global deterioration in the
fit of G-LT.

Table~\ref{tab:diag_local_g} gives local diagnostics under the G-LT
DGP. In Study~II, $\mathrm{cor}(\ell_W,\ell_L)=0.980$ and $0.989$
for $\lambda_c=0.10$ and $0.30$, respectively. LL-LT attains the
largest individual log-likelihood contribution in 33\% and 36\% of
observations on average, and exceeds G-LT locally in 40\% and 46\%.
Furthermore, $\ell_W$ correlates more strongly with
$m_{G,L}$ than with $\ell_G$ alone (0.937 vs.\ 0.892 at
$\lambda_c=0.10$; 0.966 vs.\ 0.939 at $\lambda_c=0.30$), indicating
that the best local competitor tracks $\ell_W$ more closely in the
three-model setting than G-LT does on its own. Hence, although LL-LT
is not selected globally, it enters the local comparison against W-LT
in a non-negligible fraction of observations.

\begin{table}[htb]
	\centering
	\captionsetup{justification=centering,singlelinecheck=false}
	\caption{AIC/BIC model selection proportions in the first diagnostic
	stage, with $n=250$ and $R_{\mathrm{MC}}=2{,}000$.}
	\label{tab:diag_selection}
	\renewcommand{\arraystretch}{1.12}
	\begin{tabular}{@{}llcccc@{}}
		\toprule
		Study & DGP & $\lambda_c$ & W-LT & G-LT & LL-LT \\
		\midrule
		\multirow{6}{*}{I}
		& \multirow{2}{*}{W-LT}  & $0.10$ & $0.988$ & $0.012$ & -- \\
		&                         & $0.30$ & $0.978$ & $0.022$ & -- \\
		& \multirow{2}{*}{G-LT}  & $0.10$ & $0.020$ & $0.980$ & -- \\
		&                         & $0.30$ & $0.029$ & $0.971$ & -- \\
		& \multirow{2}{*}{PW-LT} & $0.10$ & $0.001$ & $1.000$ & -- \\
		&                         & $0.30$ & $0.000$ & $1.000$ & -- \\
		\midrule
		\multirow{8}{*}{II}
		& \multirow{2}{*}{W-LT}  & $0.10$ & $0.985$ & $0.012$ & $0.003$ \\
		&                         & $0.30$ & $0.967$ & $0.021$ & $0.013$ \\
		& \multirow{2}{*}{G-LT}  & $0.10$ & $0.017$ & $0.983$ & $0.000$ \\
		&                         & $0.30$ & $0.032$ & $0.969$ & $0.000$ \\
		& \multirow{2}{*}{LL-LT} & $0.10$ & $0.015$ & $0.000$ & $0.985$ \\
		&                         & $0.30$ & $0.042$ & $0.000$ & $0.959$ \\
		& \multirow{2}{*}{PW-LT} & $0.10$ & $0.000$ & $1.000$ & $0.000$ \\
		&                         & $0.30$ & $0.000$ & $1.000$ & $0.000$ \\
		\bottomrule
	\end{tabular}
\end{table}

\begin{table}[htb]
	\centering
	\captionsetup{justification=centering,singlelinecheck=false}
	\caption{Monte Carlo averages of local correlations and dominance
	proportions under the G-LT DGP, with $n=250$ and
	$R_{\mathrm{MC}}=2{,}000$.}
	\label{tab:diag_local_g}
	\renewcommand{\arraystretch}{1.15}
	\begin{tabular}{lcccccc}
		\toprule
		Study & $\lambda_c$
		      & \makecell{$\mathrm{cor}$\\$(\ell_W,\ell_G)$}
		      & \makecell{$\mathrm{cor}$\\$(\ell_W,\ell_L)$}
		      & \makecell{$\mathrm{cor}$\\$(\ell_W,m_{G,L})$}
		      & \makecell{$\Pr$\\$(\ell_L=m)$}
		      & \makecell{$\Pr$\\$(\ell_L>\ell_G)$} \\
		\midrule
		\multirow{2}{*}{I}
		& $0.10$ & $0.893$ & -- & -- & -- & -- \\
		& $0.30$ & $0.938$ & -- & -- & -- & -- \\
		\midrule
		\multirow{2}{*}{II}
		& $0.10$ & $0.892$ & $0.980$ & $0.937$ & $0.331$ & $0.400$ \\
		& $0.30$ & $0.939$ & $0.989$ & $0.966$ & $0.361$ & $0.458$ \\
		\bottomrule
	\end{tabular}
	\smallskip\\
	\footnotesize
	Note: $m_{G,L,i}=\max\{\ell_{G,i},\ell_{L,i}\}$ and
	$m_i=\max\{\ell_{W,i},\ell_{G,i},\ell_{L,i}\}$;
	$\Pr(\ell_L=m)$ is the proportion of observations for which
	LL-LT attains the highest individual log-likelihood;
	$\Pr(\ell_L>\ell_G)$ is the proportion for which LL-LT
	exceeds G-LT locally.
\end{table}

\paragraph{Second-stage results.}
Rejection frequencies in this stage are reported as
$\Pr\left(J_M>\chi^2_{1,0.95}\right)$, where $\chi^2_{1,0.95}=3.841$ is the
95th percentile of the $\chi^2_1$ distribution.
Table~\ref{tab:J_diag_study1} summarizes Study~I. Under the W-LT DGP,
$J_{\mathrm{W}}$ has low rejection frequency while $J_{\mathrm{G}}$
rejects almost always. Under the G-LT DGP, the pattern reverses:
$J_{\mathrm{G}}$ has low rejection frequency, while $J_{\mathrm{W}}$
rejects in 0.969 and 0.937 of replications for $\lambda_c=0.10$ and
$0.30$, respectively.

Table~\ref{tab:J_diag_study2} summarizes Study~II. Under the G-LT
DGP, $J_{\mathrm{G}}$ has rejection frequencies 0.062 and 0.053,
whereas $J_{\mathrm{W}}$ rejects in 0.388 and 0.494 of replications.
Comparing the two studies, the average statistic
$\bar{J}_{\mathrm{W}}$ decreases from 14.316 to 3.959 at
$\lambda_c=0.10$, and from 11.211 to 5.494 at $\lambda_c=0.30$.
This reduction in the magnitude of
$J_{\mathrm{W}}$ mirrors the power reduction documented in
Tables~\ref{tab:study_I_J_MJ_results}
and~\ref{tab:study_II_J_MJ_results}, and is consistent with the
local-comparison mechanism identified in the first stage: in Study~II,
W-LT is compared with the best local contribution from G-LT and LL-LT
jointly, rather than with G-LT alone, and the best local competitor is
more similar to W-LT in the three-model setting than G-LT was in the
two-model setting.

This diagnostic exercise illustrates a more general
property of the \(J_m\) tests: their power depends on
how distinguishable the candidate models are from one another over the
support of the data. Adding a candidate that is locally similar to the
null model reduces the effective contrast against it, attenuating the
test's power even when the global model ranking, as captured by AIC
and BIC, remains essentially unchanged.

\begin{table}[htb]
	\centering
	\captionsetup{justification=centering,singlelinecheck=false}
	\caption{Monte Carlo averages of diagnostic quantities in Study~I,
	with $n=250$, $R_{\mathrm{MC},J}=1{,}000$, and
	$B_{\mathrm{diag}}=199$.}
	\label{tab:J_diag_study1}
	\begin{tabular}{llcccccc}
		\toprule
		DGP & $\lambda_c$
		    & $\bar{q}_{\mathrm{W}}$
		    & $\hat{\mu}_{\mathrm{W}}$
		    & $\bar{Z}_{\mathrm{W}}$
		    & $\bar{J}_{\mathrm{W}}$
		    & $\Pr\{J_{\mathrm{W}}>\chi^2_{1,0.95}\}$
		    & $\Pr\{J_{\mathrm{G}}>\chi^2_{1,0.95}\}$ \\
		\midrule
		\multirow{2}{1.8cm}{W-LT}
		& $0.10$ & $-0.035$ & $-0.035$ & $0.001$ & $1.277$ & $0.083$ & $1.000$ \\
		& $0.30$ & $-0.030$ & $-0.031$ & $0.028$ & $1.365$ & $0.087$ & $0.996$ \\
		\midrule
		\multirow{2}{1.8cm}{G-LT}
		& $0.10$ & $0.036$ & $-0.027$ & $3.673$ & $14.316$ & $0.969$ & $0.079$ \\
		& $0.30$ & $0.027$ & $-0.019$ & $3.245$ & $11.211$ & $0.937$ & $0.065$ \\
		\bottomrule
	\end{tabular}
\end{table}

\begin{table}[htb]
	\centering
	\captionsetup{justification=centering,singlelinecheck=false}
	\caption{Monte Carlo averages of \(J\)-type statistics in Study~II,
	with $n=250$, $R_{\mathrm{MC},J}=1{,}000$, and
	$B_{\mathrm{diag}}=199$.}
	\label{tab:J_diag_study2}
	\small
	\begin{tabular}{llcccccc}
		\toprule
		DGP & $\lambda_c$
		    & $\bar{J}_{\mathrm{W}}$
		    & $\Pr\{J_{\mathrm{W}}>\chi^2_{1,0.95}\}$
		    & $\bar{J}_{\mathrm{G}}$
		    & $\Pr\{J_{\mathrm{G}}>\chi^2_{1,0.95}\}$
		    & $\bar{J}_{\mathrm{L}}$
		    & $\Pr\{J_{\mathrm{L}}>\chi^2_{1,0.95}\}$ \\
		\midrule
		\multirow{2}{1.8cm}{W-LT}
		& $0.10$ & $0.531$ & $0.013$ & $32.579$ & $0.999$ & $34.095$ & $1.000$ \\
		& $0.30$ & $0.577$ & $0.013$ & $29.348$ & $0.996$ & $22.419$ & $0.998$ \\
		\midrule
		\multirow{2}{1.8cm}{G-LT}
		& $0.10$ & $3.959$ & $0.388$ & $1.122$ & $0.062$ & $56.364$ & $1.000$ \\
		& $0.30$ & $5.494$ & $0.494$ & $0.988$ & $0.053$ & $32.884$ & $1.000$ \\
		\midrule
		\multirow{2}{1.8cm}{LL-LT}
		& $0.10$ & $9.058$ & $0.787$ & $86.831$ & $1.000$ & $1.865$ & $0.139$ \\
		& $0.30$ & $6.928$ & $0.632$ & $68.812$ & $1.000$ & $1.385$ & $0.101$ \\
		\bottomrule
	\end{tabular}
\end{table}

\section{Empirical Application}
\label{sec:application}

This section illustrates the proposed procedure on a
real dataset, with emphasis on the complementary information provided
by the \(MJ\) test relative to AIC and BIC. The key feature of the
application is that the three candidate models rank identically under
both information criteria in both analysis groups, yet the \(MJ\) test
reaches different conclusions about the adequacy of the candidate set
in the two groups, revealing a distinction that AIC and BIC cannot
make.

We consider data from the Hospital Cancer Registry of the Funda\c{c}\~ao
Oncocentro de S\~ao Paulo (RHC/FOSP) \cite{fosp_rhc}, a public cancer
registry database coordinated by FOSP in the state of S\~ao Paulo,
Brazil. The dataset comprises $12{,}030$ women diagnosed with cervical
cancer (ICD-10 C53) between 2012 and 2022. The response is survival
time, expressed in years, and the event of interest
is death from cervical cancer. Patients who did not experience the event
during follow-up are treated as right-censored. The analysis is
stratified by chemotherapy record, denoted by $\mathrm{CHEMO}=0$ (no
record) and $\mathrm{CHEMO}=1$ (record present). This variable is used
only to define two groups; no causal interpretation is intended, since
treatment registration may be associated with disease stage, severity,
therapeutic indication, and other clinical factors not controlled for
here.

\subsection{Descriptive analysis}

The group without chemotherapy record contains $4{,}880$ patients, with
$1{,}475$ events and an observed event proportion of $0.302$, whereas
the group with chemotherapy record contains $7{,}150$ patients, with
$2{,}702$ events and an observed event proportion of $0.378$. The
Kaplan--Meier estimates reveal distinct survival patterns between the
two groups. Survival is higher early in the follow-up for the group
with chemotherapy record: the estimates at $0.5$ and $1$ year are
$0.960$ and $0.857$, respectively, compared with $0.847$ and $0.774$
for the group without chemotherapy record. This pattern reverses at
longer times: at $3$, $5$, and $8$ years, the estimated survival
probabilities are $0.600$, $0.512$, and $0.466$ for the group with
chemotherapy record, versus $0.675$, $0.647$, and $0.623$ for the
group without. The crossing of the Kaplan--Meier
curves suggests that the two groups have different risk dynamics over
time, which may affect the adequacy of standard parametric LT models.

\subsection{Fitted models and information criteria}

For each chemotherapy group, the Weibull-LT, Gompertz-LT, and
log-logistic-LT models were fitted separately, denoted by W-LT, G-LT,
and LL-LT, respectively. Table~\ref{tab:LT_parameter_estimates} presents
parameter estimates and standard errors for all three models in both
groups. Standard errors are derived from the observed information matrix,
computed as the negative of the analytic Hessian of the log-likelihood
evaluated at the MLE. The parameter $\widehat{p}_0$
denotes the estimated cure fraction, i.e., the estimated proportion of
individuals not expected to experience the event under the fitted model.
Across all three fitted LT models, $\widehat{p}_0$ is consistently lower
in the group with chemotherapy record. The estimates are $0.634$, $0.637$,
and $0.610$ for W-LT, G-LT, and LL-LT when $\mathrm{CHEMO}=0$, and
$0.477$, $0.467$, and $0.453$ when $\mathrm{CHEMO}=1$, consistent with
the lower long-term survival observed in the Kaplan--Meier estimates for
that group.

\begin{table}[htb]
  \centering
  \caption{Parameter estimates and standard errors for the fitted LT
           models. For W-LT and LL-LT, $(\theta_1,\theta_2)=(k,\lambda)$;
           for G-LT, $(\theta_1,\theta_2)=(\alpha,\xi)$.}
  \label{tab:LT_parameter_estimates}
  \begin{tabular}{llcccc}
    \toprule
    CHEMO & Model & Quantity & $\widehat{\theta}_1$ & $\widehat{\theta}_2$ & $\widehat{p}_0$ \\
    \midrule
    \multirow{6}{*}{$0$}
    & \multirow{2}{*}{W-LT}  & Estimate & $0.903$ & $0.903$ & $0.634$ \\
    &                         & SE       & $0.020$ & $0.031$ & $0.008$ \\
    & \multirow{2}{*}{G-LT}  & Estimate & $0.898$ & $<0.001$ & $0.637$ \\
    &                         & SE       & $0.029$ & $0.001$ & $0.008$ \\
    & \multirow{2}{*}{LL-LT} & Estimate & $1.202$ & $1.438$ & $0.610$ \\
    &                         & SE       & $0.034$ & $0.092$ & $0.010$ \\
    \midrule
    \multirow{6}{*}{$1$}
    & \multirow{2}{*}{W-LT}  & Estimate & $1.450$ & $0.300$ & $0.477$ \\
    &                         & SE       & $0.023$ & $0.009$ & $0.008$ \\
    & \multirow{2}{*}{G-LT}  & Estimate & $0.314$ & $0.198$ & $0.467$ \\
    &                         & SE       & $0.010$ & $0.017$ & $0.009$ \\
    & \multirow{2}{*}{LL-LT} & Estimate & $1.977$ & $0.342$ & $0.453$ \\
    &                         & SE       & $0.041$ & $0.013$ & $0.009$ \\
    \bottomrule
  \end{tabular}
\end{table}

As shown in Table~\ref{tab:LT_ic}, LL-LT yields the lowest AIC and BIC
values in both chemotherapy groups. According to both information
criteria, LL-LT is therefore the preferred model within the original
LT candidate set in both groups.

\begin{table}[ht]
  \centering
  \caption{Information criteria for the fitted W-LT, G-LT, and LL-LT
           models by chemotherapy group.}
  \label{tab:LT_ic}
  \begin{tabular}{llrr}
    \toprule
    CHEMO & Model & AIC & BIC \\
    \midrule
    \multirow{3}{*}{$0$}
    & W-LT  & $8{,}099.827$  & $8{,}119.306$ \\
    & G-LT  & $8{,}122.800$  & $8{,}142.279$ \\
    & LL-LT & $8{,}022.159$  & $8{,}041.638$ \\
    \midrule
    \multirow{3}{*}{$1$}
    & W-LT  & $15{,}312.350$ & $15{,}332.975$ \\
    & G-LT  & $15{,}626.112$ & $15{,}646.737$ \\
    & LL-LT & $15{,}115.700$ & $15{,}136.325$ \\
    \bottomrule
  \end{tabular}
\end{table}

\subsection{Non-nested tests and graphical assessment}

Since AIC and BIC provide only relative comparisons among candidate
models, the bootstrap \(MJ\) procedure was applied to assess whether the
original LT candidate set provides an adequate description of each
chemotherapy group, using $B=499$ bootstrap replications. For
$\mathrm{CHEMO}=0$, the \(MJ\) procedure did not reject the candidate
set ($p$-value $=0.230$), and LL-LT was selected as the preferred
model, in agreement with AIC and BIC. For
$\mathrm{CHEMO}=1$, however, the \(MJ\) procedure rejected the candidate
set ($p$-value $=0.002$). This means that, although
LL-LT is the best model in a relative sense within the original candidate
set for both groups, it cannot be regarded as an adequate description
of the survival pattern for the group with chemotherapy record in an
absolute sense. This is the key distinction that AIC and BIC cannot
provide: both criteria selected LL-LT in both groups without signaling
any inadequacy.

\begin{figure}[ht]
  \centering
  \includegraphics[width=0.75\linewidth]{%
    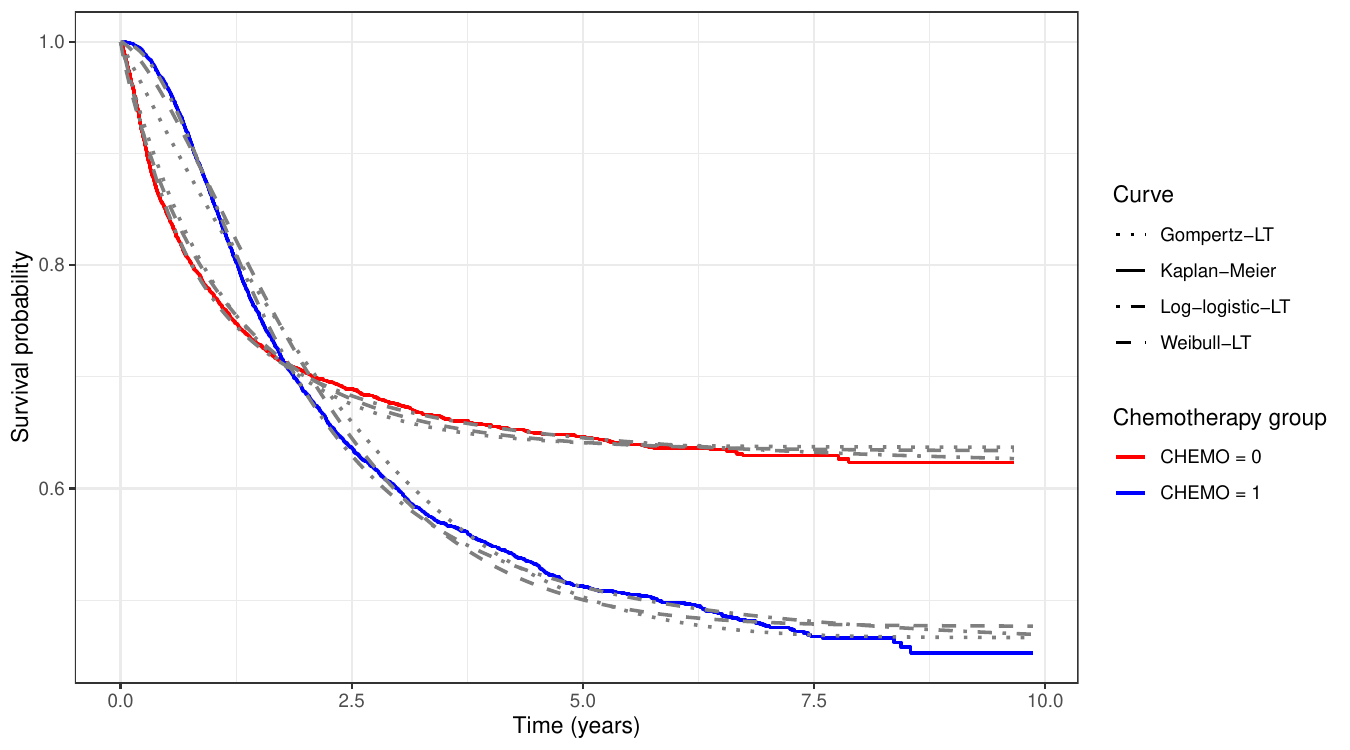}
  \caption{Kaplan--Meier estimates and fitted survival curves from the
           W-LT, G-LT, and LL-LT models by chemotherapy group.}
  \label{fig:LT_fitted_chemotherapy}
\end{figure}

\begin{figure}[ht]
  \centering
  \includegraphics[width=0.75\linewidth]{%
    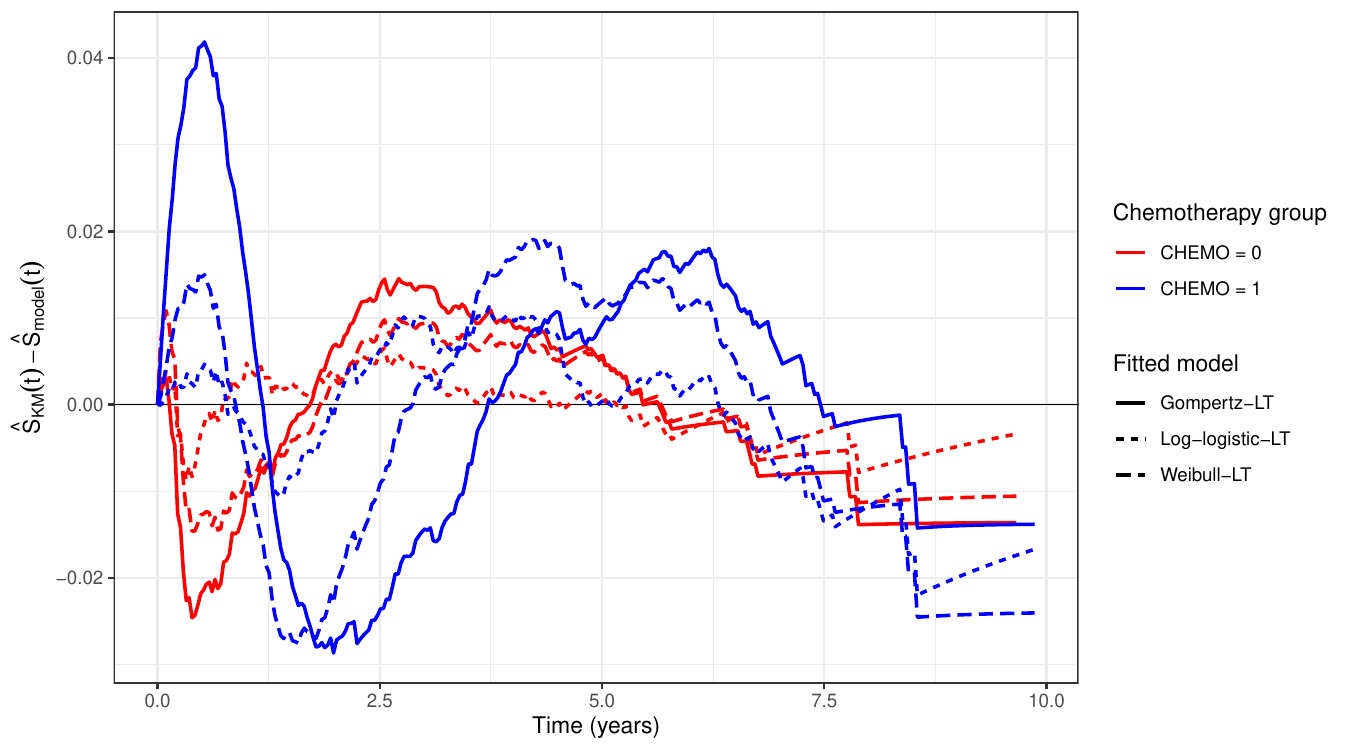}
  \caption{Pointwise differences between the Kaplan--Meier estimates
           and the fitted LT survival curves by chemotherapy group.}
  \label{fig:KM_minus_LT_chemotherapy}
\end{figure}

The graphical comparison in Figures~\ref{fig:LT_fitted_chemotherapy}
and~\ref{fig:KM_minus_LT_chemotherapy} corroborates this distinction.
The fitted W-LT, G-LT, and LL-LT survival curves follow the
Kaplan--Meier estimates reasonably closely in both groups, but the
agreement is better for $\mathrm{CHEMO}=0$. This is more clearly seen
in the pointwise differences in
Figure~\ref{fig:KM_minus_LT_chemotherapy}, computed as the
Kaplan--Meier estimate minus the fitted survival curve. For LL-LT, the
maximum absolute difference is $0.009$ in $\mathrm{CHEMO}=0$ and
$0.022$ in $\mathrm{CHEMO}=1$, with mean absolute differences of
$0.003$ and $0.008$, respectively. Although numerically
modest, the discrepancies in $\mathrm{CHEMO}=1$ are more systematic
across the follow-up period, consistent with the rejection by the \(MJ\)
test.

\subsection{Complementary analysis with more flexible models}

The rejection of the original LT candidate set for
$\mathrm{CHEMO}=1$ motivates a search for more flexible specifications.
We fitted additional LT models based on the Generalized~F and
Generalized gamma distributions, as well as a Piecewise-LT model,
denoted by GF-LT, GG-LT, and PW-LT, respectively. Since LL-LT was the
preferred model within the original candidate set, it serves as the
reference specification in this comparison.

As shown in Table~\ref{tab:flexible_cure}, GF-LT yields the smallest
AIC in both groups. For $\mathrm{CHEMO}=0$, the gain is modest and BIC
still favors LL-LT, suggesting that the additional flexibility of GF-LT
is not strongly supported by the data in that group. For
$\mathrm{CHEMO}=1$, the evidence is more consistently favorable to
GF-LT: both AIC ($15{,}086.905$ versus $15{,}115.700$) and BIC
($15{,}121.280$ versus $15{,}136.325$) favor the more flexible model.

\begin{table}[ht]
  \centering
  \caption{Information criteria for complementary LT specifications by
           chemotherapy group.}
  \label{tab:flexible_cure}
  \begin{tabular}{llrr}
    \toprule
    CHEMO & Model & AIC & BIC \\
    \midrule
    \multirow{4}{*}{$0$}
    & GF-LT & $8{,}015.816$  & $8{,}048.280$ \\
    & GG-LT & $8{,}022.140$  & $8{,}048.111$ \\
    & LL-LT & $8{,}022.159$  & $8{,}041.638$ \\
    & PW-LT & $8{,}055.444$  & $8{,}081.416$ \\
    \midrule
    \multirow{4}{*}{$1$}
    & GF-LT & $15{,}086.905$ & $15{,}121.280$ \\
    & LL-LT & $15{,}115.700$ & $15{,}136.325$ \\
    & GG-LT & $15{,}117.190$ & $15{,}144.690$ \\
    & PW-LT & $15{,}264.200$ & $15{,}291.699$ \\
    \bottomrule
  \end{tabular}
\end{table}

A likelihood-ratio test comparing LL-LT and GF-LT provides a focused
assessment of the additional flexibility of the Generalized~F
distribution. For $\mathrm{CHEMO}=0$, the test is significant
($\mathrm{LR}=10.343$, $\mathrm{df}=2$, $p=0.006$), but BIC still
favors LL-LT, so the overall interpretation for this group does not
change. For $\mathrm{CHEMO}=1$, the evidence is stronger
($\mathrm{LR}=32.795$, $\mathrm{df}=2$, $p<0.001$), and both AIC and
BIC favor GF-LT, indicating that the additional flexibility is
genuinely warranted in this group.

\begin{figure}[ht]
  \centering
  \includegraphics[width=0.75\linewidth]{%
    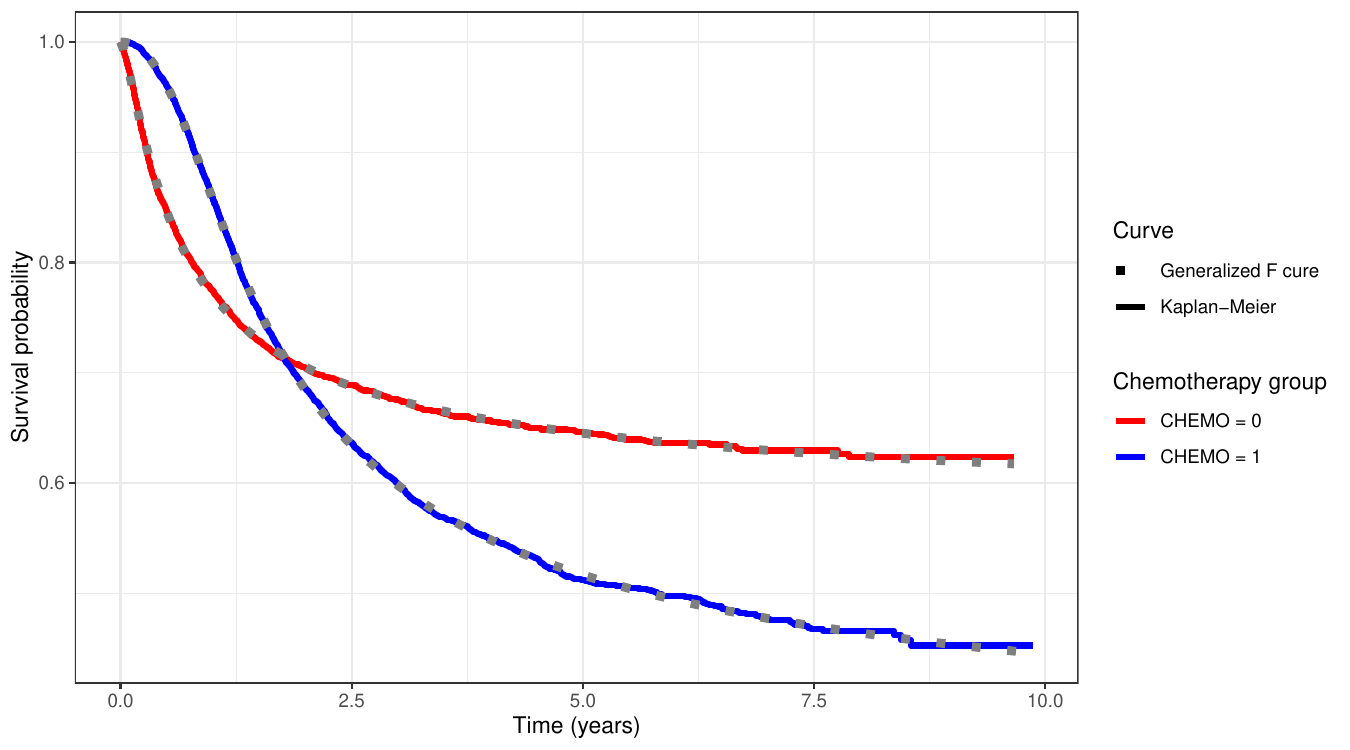}
  \caption{Kaplan--Meier estimates and fitted survival curves from the
           Generalized F-LT model by chemotherapy group.}
  \label{fig:GF_cure_chemotherapy}
\end{figure}

Figure~\ref{fig:GF_cure_chemotherapy} shows that the GF-LT model
provides close visual agreement with the Kaplan--Meier estimates in
both groups. Together, these results confirm the
interpretation suggested by the \(MJ\) test: for $\mathrm{CHEMO}=0$,
LL-LT provides an adequate and parsimonious working description; for
$\mathrm{CHEMO}=1$, the original LT candidate set is inadequate, and a
more flexible specification --- here, GF-LT --- is needed to capture
the observed survival pattern. The \(MJ\) test thus signaled a genuine
distributional inadequacy in one group that AIC and BIC, by design,
could not detect.

\section{Concluding Remarks}
\label{sec:conclusion}

We have developed a score-based framework for testing and selecting
among non-nested parametric survival models with a cure fraction.
The central methodological contribution is a new
augmentation strategy: rather than adding the fitted linear predictor
of the competing model as an extra regressor, as in the classical \(J\)
test of Davidson and
MacKinnon~\cite{davidson1981} and its GAMLSS extension by Cribari-Neto and
Lucena~\cite{cribari2017},
we augment the log-likelihood of the null model with the individual
log-likelihood differences evaluated at the competing MLEs. A score
test then assesses whether this additional information is redundant.
Unlike the method of Cribari-Neto and Lucena~ \cite{cribari2017}, the framework does not
require a regression structure and remains valid when the candidate
models differ solely in their postulated baseline distribution ---
which is the primary setting of interest in cure-fraction survival
analysis.

The framework has three features that distinguish it from the Vuong test~\cite{vuong1989}. First, we test the specific null hypothesis
that a given model is the true data-generating process, which has a
direct substantive interpretation and avoids the size distortion that
the Vuong test incurs when applied to this question. Second, by working
with an unsigned test statistic and estimating the Kullback-Leibler
bias via parametric bootstrap, the procedure extends naturally to
\(M\geq 2\) candidate models. Third, the \(MJ\) global test provides a
principled way to detect that the entire candidate set is inadequate
--- a possibility that information criteria cannot entertain.
A practical computational advantage is that a single
set of bootstrap replicates, generated under the model assumed correct
by each null hypothesis, simultaneously estimates the KL bias and
calibrates the critical values, so the bias correction incurs no
additional simulation cost.

The Monte Carlo evidence confirms the satisfactory
finite-sample behavior of the proposed tests under full-bootstrap
calibration. The individual \(J_m\) tests control size close to the
nominal level when the null model is correctly specified, and their
power increases with sample size. The \(MJ\) test rejects rarely when
at least one candidate model is correctly specified, but rejects with
high frequency when all candidates are misspecified, as in the
piecewise-exponential DGP scenarios. The diagnostic analysis in
Section~\ref{sec:simulation} further shows that adding a third
candidate model to the fitted set can reduce the power of individual
\(J_m\) tests when the new candidate is locally similar to the null model
--- a consequence of the max-competitor construction used to define the
log-likelihood differences for $M>2$.

The empirical application to cervical cancer survival data illustrates
the practical value of the \(MJ\) test. In one of the two analysis groups,
the test rejected all three candidate models despite the Log-logistic-LT
being selected by both AIC and BIC, demonstrating that
relative model selection and absolute model adequacy are distinct
questions that require different tools. The complementary analysis
confirmed that a more flexible parametric family was needed to capture
the observed survival pattern in that group.

Several extensions of the proposed framework are
worth pursuing. In many survival studies, the primary goal is
regression: the cure fraction and the distributional parameters of
the susceptible subpopulation are linked to covariates through
regression submodels. Extending the stimulated log-likelihood approach
to this setting is conceptually straightforward --- the augmentation
acts at the level of the full log-likelihood, not the individual
submodels, so the score statistic retains the same form --- but the
finite-sample properties of the bootstrap bias estimator under a
regression structure merit separate Monte Carlo investigation. A
further extension is to settings with interval censoring or
competing risks, where the individual log-likelihood contributions
take different forms but the general augmentation strategy remains
applicable. We leave these directions for future work.

\section*{Acknowledgements}

The authors gratefully acknowledge the computational resources provided by the Centro de Computa\c{c}\~ao de Alto Desempenho at the Federal University of Pernambuco, on the Boi Voador cluster.

\section*{Conflict of interest}

The author(s) declared no potential conflicts of interest with respect to the
research, authorship, and/or publication of this article.

\section*{Funding}
This work was supported by the Coordenação de Aperfeiçoamento de Pessoal de
Nível Superior (CAPES) through a postdoctoral fellowship awarded to Cynthia
A. V. Tojeiro under the PDPG-FAP III program (process number
88887.235722/2025-00).

Francisco Cribari-Neto acknowledges financial support from the Conselho Nacional
de Desenvolvimento Científico e Tecnológico (CNPq), grant 304646/2023-7.

\section*{Data availability statement}

The cervical cancer dataset analyzed in this study is publicly available from
the Fundação Oncocentro de São Paulo (FOSP), through the RHC/FOSP database. The
real cancer registry datasets are not redistributed by the authors.


\bibliographystyle{plainnat}
\bibliography{referencias}

\end{document}